\documentstyle[12pt]{article}
\newcommand{\nc}{\newcommand}

\nc{\dbar}{\bar{\partial}}
\nc{\be}{\begin{equation}}
\nc{\ee}{\end{equation}}

\catcode`\@=11

\@addtoreset{equation}{section}

\def\@normalsize{\@setsize\normalsize{15pt}\xiipt\@xiipt
\abovedisplayskip 14pt plus3pt minus3pt%
\belowdisplayskip \abovedisplayskip
\abovedisplayshortskip  \z@ plus3pt%
\belowdisplayshortskip  7pt plus3.5pt minus0pt}
\def\small{\@setsize\small{13.6pt}\xipt\@xipt
\abovedisplayskip 13pt plus3pt minus3pt%
\belowdisplayskip \abovedisplayskip
\abovedisplayshortskip  \z@ plus3pt%
\belowdisplayshortskip  7pt plus3.5pt minus0pt
\def\@listi{\parsep 4.5pt plus 2pt minus 1pt
            \itemsep \parsep
            \topsep 9pt plus 3pt minus 3pt}}

\def\underline#1{\relax\ifmmode\@@underline#1\else
        $\@@underline{\hbox{#1}}$\relax\fi}
\@twosidetrue
\relax

\catcode`@=12

\catcode`\@=11

\def\section{\@startsection{section}{1}{\z@}{3.5ex plus 1ex minus
   .2ex}{2.3ex plus .2ex}{\large\bf}}

\begin{document}
\pagestyle{empty}
\rightline{CPTH-RR370.0895}
\rightline{hep-th/9509009}
\begin{center}
{\Large \bf Exact Monodromy Group of N=2 Heterotic Superstring} \\
\vspace{1.5 cm}
{\large I. Antoniadis and H. Partouche} \\
\vspace{5 mm}
{\em 	Centre de Physique Th\'{e}orique\footnote{\sl Laboratoire Propre du CNRS
UPR A.0014.} \\
			Ecole Polytechnique \\
			F-91128 Palaiseau cedex, France} \\
\vspace{1.5 cm}
{\bf Abstract}
\end{center}
We describe an $N=2$ heterotic superstring model of rank-3 which is dual
to the type-II string compactified on a Calabi-Yau manifold with Betti
numbers $b_{1,1}=2$ and $b_{1,2}=86$. We show that the exact duality symmetry
found from the type II realization contains the perturbative duality group of
the heterotic model, as well as the exact quantum monodromies of the rigid
$SU(2)$ super-Yang-Mills theory. Moreover, it contains a non-perturbative
monodromy which is stringy in origin and corresponds roughly to an exchange of
the string coupling with the compactification radius.
\vspace{1.5 cm}\\
September 1995

\newpage
\setcounter{footnote}{0}
\setcounter{page}{1}
\pagestyle{plain}
\section{Introduction}

During the last year, there has been considerable progress in understanding
non-\break perturbative superstring theory based on the conjectured
string-string
duality \cite{{s-s}}. $N=2$ supersymmetric compactifications provide the
first examples of non-trivial dynamics which are simple enough to be treated as
laboratories for studying non-perturbative effects. The main observation is
that
the dilaton which plays the role of the string coupling constant belongs to a
vector multiplet in the heterotic string and to a hypermultiplet in the type II
string. Since abelian vector multiplets and neutral hypermultiplets do not
interact in the low energy theory, tree-level type II theory provides the exact
answer for the description of the moduli space of vector multiplets while
tree-level heterotic theory is exact for hypermultiplets.

An interesting class of dual pairs consists in heterotic compactifications on
$K_3\times T^2$, on the one side, with type II compactifications on Calabi-Yau
threefolds, on the other side \cite{{Vafa},{FHSV}}. At generic points of their
moduli space, the gauge group is abelian $U(1)^r$ and there are no charged
massless
hypermultiplets. In terms of the two Betti numbers of the Calabi-Yau manifold,
$b_{1,1}$ and $b_{1,2}$, the rank is $r=b_{1,1}+1$ and the number of
hypermultiplets $b_{1,2}+1$ where the $+1$ count the graviphoton and the type
II
dilaton, respectively. Moreover one considers only models with $b_{1,1}\ge 1$
to
count for the heterotic dilaton. Among several concrete examples of dual pairs
which have been proposed in ref. \cite{Vafa}, there have been more quantitative
tests only for the vector multiplet sector of two models of rank 3 and 4 which
correspond in the heterotic side to models with 1 and 2 moduli, respectively,
besides the dilaton \cite{{Vafa},{Louis},{Klemm.95},{W2g}}. Their exact quantum
moduli space is described on the type II side by the geometry of the Calabi-Yau
hypersurfaces of degree 12 and 24, $X_{12}$ and $X_{24}$, correspondingly,
which
were studied extensively in refs. \cite{{Candelas},{Klemm.93}} using mirror
symmetry.

The main test consists in computing the perturbative correction to the
heterotic
prepotential \cite{{yuka},{AFGNT},{deWit}} which, due to $N=2$
non-renormalization
theorems, occurs only at the one loop level, and compare it with the exact
expression obtained from type II \cite{{Vafa},{Louis},{Klemm.95},{W2g}}. This
procedure requires the correct identification of the heterotic dilaton as one
of
the $b_{1,1}$ moduli of type II which allows to take the weak coupling limit
corresponding to the large complex structure in the mirror Calabi-Yau. Besides
this comparison of the low energy theory, further non-trivial tests have been
performed for the rank 3 model by analyzing the structure of higher dimensional
interactions \cite{W2g}.

Given these perturbative tests which strongly support the proposed equivalence
between heterotic/type II dual pairs, it would be very interesting and
instructive to study, using duality, the structure of non-perturbative
effects in
heterotic $N=2$ compactifications. A technical difficulty is that the exact
results based on type II become simple only in certain variables parameterizing
the complex structure of the mirror manifold, while the expressions for the
periods and the mirror map to the special coordinates appropriate for the
description of the heterotic vacua, are known only by their instanton
expansions. A possible way to bypass this difficulty is to study the exact
quantum symmetry and the effects of non-perturbative corrections to the
perturbative $T$-duality transformations. In this work we address these
issues in the context of a different rank-3 dual pair example based, on the
type
II side, on the Calabi-Yau hypersurface of degree 8, $X_{8}$, for which the
exact monodromy group was worked out in detail in ref. \cite{Candelas}.

In section 2, we review the type II model $X_{8}$ and its large complex
structure
limit which motivates its dual interpretation by identifying the heterotic
dilaton
and the classical $T$-duality group. The later is actually ${\Gamma_0(2)}_+$,
as
was claimed in ref. \cite{Klemm.95}, which is generated by the transformations
$T\rightarrow T+1$ and $T\rightarrow -1/2T$ and shares very similar
properties with
$SL(2,{\bf Z})$.

In Section 3, we describe the heterotic dual
realization and
compute the one loop correction to the prepotential. At the particular
point of the
$T$-modulus space $T=i/{\sqrt 2}$ which is an order 2 fixed point, the
$U(1)$ of $T$ is extended to $SU(2)$ because of the appearance of two extra
charged massless vector multiplets, and the prepotential develops a logarithmic
singularity. The main difference with the other rank-3 dual pair based on
$X_{12}$, besides the $T$-duality group, is that the $SU(2)$ enhanced gauge
symmetry appears in this case with Kac-Moody level 2. Comparing the heterotic
perturbative prepotential with the exact answer from the type II side in the
weak coupling limit, we find a
perfect agreement of the two expressions.

In Section 4, we derive the quantum monodromies of the one loop prepotential
associated to its logarithmic singularity, following the method of ref.
\cite{AFGNT}. As a result the classical duality group gets modified to a
subgroup of $Sp(6,{\bf Z})$ depending on 8 integer parameters. One of its two
generators corresponding to the transformation $T\rightarrow -1/2T$ is
identified with the Weyl reflection of the $SU(2)$ enhanced symmetry with its
quantum monodromy determined in the rigid supersymmetric theory
\cite{Seiberg-Witten}.

In Section 5, we present the exact symmetry group determined from type II using
the results of ref. \cite{Candelas}, and we identify the generators of the
perturbative $T$-duality group and the quantized axionic shift which form a
subgroup. Moreover among the remaining elements, we identify the
non-perturbative quantum monodromy of the rigid $SU(2)$ theory associated to
the
appearance of massless monopoles or dyons \cite{Seiberg-Witten}. This is a
non-perturbative test of string-string duality which, in particular, exchanges
world-sheet instantons of type II string with spacetime instantons of heterotic
string. It also confirms the expectation that the enhanced $SU(2)$ symmetric
point which is present in perturbation theory of $N=2$ heterotic
compactifications, disappears in the exact quantum theory, and the
corresponding
$SU(2)$ singularity splits into two branches where non-perturbative solitonic
states become massless. It is remarkable that under string duality, this
non-perturbative splitting is mapped on the type II side into the conifold
locus
of Calabi-Yau manifolds, while the dyonic hypermultiplets should be mapped into
charged black holes which become massless at the conifold locus \cite{Stro}.
The
exact symmetry group contains an additional non-perturbative monodromy which is
stringy in origin and is related to the dilaton. It corresponds to an
$S\leftrightarrow T$ duality \cite{Klemm.95} and implies the existence of new
massless solitons with charges under the extra $U(1)$'s. All these properties
seem to be much more generic than for this particular example.\footnote{After
the completion of this work, we received the paper of ref.
\cite{KKLMV} where similar results were obtained for the rank-3 dual pair based
on $X_{12}$.}

\section{The type II model}

Let us consider the rank-3 type II superstring compactification on the manifold
$X_{8}$ with Betti numbers $b_{1,1}=2$ and $b_{1,2}=86$, giving
rise to 2 vector multiplets (besides the graviphoton) and $86+1$
hypermultiplets
including the dilaton. This manifold is defined as a hypersurface of degree 8
in
the weighted projective space ${\bf WCP^{4}}[1,1,2,2,2]$. The
moduli space of the
two vector multiplets can be studied using mirror symmetry
\cite{Candelas}. The mirror
$X_{8}^{*}$
may be identified with the family of Calabi-Yau threefolds of the form $\{
{\cal
P} = 0 \}/{\bf Z}_{4}^{3}$, where
\begin{equation}
{\cal P} = z_{1}^{8}+z_{2}^{8}+z_{3}^{4}+z_{4}^{4}+z_{5}^{4}-8 \psi
z_{1} z_{2} z_{3} z_{4} z_{5} -2\phi z_{1}^{4} z_{2}^{4}\ ,
\label{X8}
\end{equation}
depending on two complex structure deformation parameters $\psi$ and $\phi$.
The three discrete group factors ${\bf Z}_{4}$ act on the coordinates as
$(z_2, z_{2+m})\rightarrow (-iz_2, iz_{2+m})$ for $m=1,2,3$, respectively.

For a good description of the moduli space, it is convenient to enlarge
${\bf Z}_{4}^{3}$ to a group $\widehat{G}$ acting also on $\psi$ and $\phi$,
\begin{equation}
(z_{1}, z_{2}, z_{3}, z_{4}, z_{5};\psi,\phi)\rightarrow(\alpha^{a_{1}}z_{1},
\alpha^{a_{2}}z_{2},\alpha^{2a_{3}} z_{3},\alpha^{2a_{4}} z_{4},\alpha^{2a_{5}}
z_{5};\alpha^{-a}\psi,\alpha^{-4a}\phi),
\end{equation}
where $\alpha=e^{2i\pi /8}$, $a_{i}$ are integers and
$a=a_{1}+a_{2}+2a_{3}+2a_{4}+2a_{5}$. $X_{8}^{*}$ is then obtained by
considering the quotient $\{ {\cal P} = 0 \}/\widehat{G}$ and mod out the
parameter space  by a ${\bf Z}_8$ with generator
\begin{equation}
(\psi,\phi)\rightarrow (\alpha\psi,-\phi).
\label{ordre 8}
\end{equation}
To describe the large complex structure limit, it is also convenient to
introduce the variables
\begin{equation}
x=-\frac{2 \phi}{(8 \psi)^{4}} \quad \mbox{ and }\quad y =
\frac{1}{(2\phi)^{2}}\ .
\end{equation}

The Yukawa couplings are worked out in an explicit form in refs.
\cite{{Candelas},{Klemm.93}}. As follows from $N=2$ special geometry, there
is a preferred coordinate system in which they are given as triple derivatives
of an analytic prepotential \cite{{sugra},{special}}. The corresponding special
coordinates
$t_{1}$ and $t_{2}$ are given by the inverted mirror map,
\begin{eqnarray}
x(t_{1},t_{2})& =& \frac{1}{h(t_{1})}( 1+ {\cal O}_{t_{1}}(q_{_2}))
\label{mirror map} \\
y(t_{1},t_{2})& =& q_{_2} l(t_{1}) ( 1+ {\cal O}_{t_{1}}(q_{_2})) \nonumber
\end{eqnarray}
where $q_{_1}=e^{2i \pi t_{1}}$,  $q_{_2}=e^{2i \pi t_{2}}$ and
${\cal O}_{t_{1}}(q_{_2})$ denotes terms which are at least of order one in
$q_{_2}$  and order zero in $q_{_1}$. The function $l(t_{1})=1+{\cal
O}(q_{_1})$,
while $h(t_{1})=1/q_{_1}+104+{\cal O}(q_{_1})$ will be given below. In these
coordinates, the Yukawa couplings ${\cal Y}_{ijk}$ are derived by a
prepotential
${\cal Y}$,
\begin{equation}
{\cal Y}_{ijk}\equiv -\partial_{t_{i}}\partial_{t_{j}}\partial_{t_{k}}{\cal Y},
\end{equation}
and they are given as power series in $q_1$ and $q_2$:
\begin{eqnarray}
{\cal Y}_{111} &=& 8 + \sum_{n\ge 1}y_n q_{_1}^n+ {\cal O}_{t_{1}}(q_{_2})
\nonumber \\
{\cal Y}_{112} &=& 4+ {\cal O}_{t_{1}}(q_{_2})\label{yukawa}\\
{\cal Y}_{122} &=& {\cal O}_{t_{1}}(q_{_2}) \nonumber \\
{\cal Y}_{222} &=& {\cal O}_{t_{1}}(q_{_2}) \nonumber
\end{eqnarray}
where $y_n$ are known integer coefficients related to the world-sheet instanton
numbers \cite{Candelas}.

{}From eq.(\ref{yukawa}) one can extract the prepotential
\begin{eqnarray}
{\cal Y} &=& -2t_{1}^{2}t_{2} -\frac{4}{3}t_{1}^{3}+ \frac{\alpha}{2}
t_{1}^{2}+(\delta-\frac{2}{3})t_{1}+\xi + g(t_{1}) \label{crochet} \\
  & & + \beta t_{1}t_{2}+\frac{\gamma}{2} t_{2}^{2}+\epsilon t_{2} +
       g^{^{NP}}(t_{1},t_{2}) \nonumber
\end{eqnarray}
where $\alpha$, $\beta$, $\gamma$, $\delta$, $\epsilon$ and $\xi$ are arbitrary
integration constants, and
\begin{eqnarray}
g(t_{1}) &=&{1\over 8i\pi^3}\sum_{n \geq 1}{y_n\over n^3}q_{_1}^{n}
\label{gdef}\\
g^{^{NP}}(t_{1},t_{2}) &=& \sum_{n \geq 1, m \geq 0}g^{^{NP}}_{nm}
q_{_2}^{n} q_{_1}^{m}\nonumber
\end{eqnarray}
The above form of the prepotential suggests that this type II model has a
heterotic dual realization by identifying $t_2$ with the heterotic
dilaton $S$ and $t_1$ with the single heterotic modulus $T$ \cite{Vafa}. In
fact,
in this case, the general heterotic prepotential $F$ takes the form
\begin{equation}
{ F} = -2ST^{2} + f(T) + f^{^{NP}}(T,S)\ ,
\label{prep}
\end{equation}
where the first two terms of the right hand side correspond to the classical
and
one-loop contributions, respectively. $f^{^{NP}}$
denotes the non perturbative corrections
\begin{equation}
f^{^{NP}}(T,S) = \sum_{n \geq 1}f^{^{NP}}_{n}(T)
q_{_S}^{n}
\end{equation}
where $q_{_S}=e^{2i\pi S}$ and $\langle
S\rangle =\frac{\theta}{\pi}+i \frac{8\pi}{g_s^2}$ in terms of the
$\theta$-angle and the four-dimensional string coupling constant $g_s$.

This identification is motivated by the fact that in the weak coupling limit
$q_{_2}\rightarrow 0$ (or equivalently $y\rightarrow 0$), the discriminant
locus
of the mirror Calabi-Yau $X_{8}^{*}$
\begin{equation}
\label{conifold}
\Delta=\{ (1-2^{8}x)^{2}-2^{18}x^2y\}\{ 1-4y\}=0
\end{equation}
becomes a perfect square, in complete analogy with the situation in the
rigid $SU(2)$ $N=2$ supersymmetric Yang-Mills theory where the classically
singular $SU(2)$ point splits into two separate branches in the quantum theory
\cite{Seiberg-Witten}. The first factor in eq.(\ref{conifold}) defines
the conifold singularity while the second factor gives rise to an isolated
singularity $y=1/4$ corresponding to the infinite strong coupling limit
$S\rightarrow 0$ in the heterotic theory \cite{{Candelas},{Vafa}}.

The guide to construct the heterotic realization is to first find the classical
$T$-duality group \cite{{Klemm.95},{Yau}}. This can be determined by the set of
transformations which identify all solutions of the conifold
singularity in the weak coupling limit $y=0,x=1/256$ in terms of the special
coordinate $T$,
\begin{equation}
h(T)=256 \quad\mbox{ and }\quad S=i\infty\ ,
\label{h=256}
\end{equation}
where $h$ is defined in eq. (\ref{mirror map}). In fact $h$ was found to be (up
to an additive constant 104) the Haupmodul for ${\Gamma_{0}(2)}_+$ which should
play the role of the heterotic $T$-duality group,
\begin{equation}
h(T) = \frac{\left[ \, \theta_{3}^{4}(T)+\theta_{4}^{4}(T) \,
\right]^{4}}{16\left[ \,  \eta(T)\eta(2T) \, \right]^{8}}\ ,
\label{hfun}
\end{equation}
where $\theta_{3,4}$ are the Jacobi $\theta$-functions and $\eta$ the Dedekind
function.

${\Gamma_{0}(2)}_+$ is the group $\Gamma_{0}(2)$ of modular transformations
\begin{equation}
T \rightarrow \frac{aT+b}{cT+d} \equiv MT\quad ;\quad M=
\left( \begin{array}{cc}
         a & b \\
         c & d
       \end{array} \right)
\label{Mt}
\end{equation}
with $a,b,c,d$ integers, $c$ even and $\det M =1$, together with an
Atkin-Lenhner involution $W$ of determinant two \cite{Conway},
\begin{equation}
W :\quad T\rightarrow -\frac{1}{2T} \ .
\label{W}
\end{equation}
The full group can be generated in terms of two elements which, for instance,
can be chosen to be $W$ and the translation
\begin{equation}
U :\quad T \rightarrow T+1 \ ,
\label{U}
\end{equation}
or equivalently,
\begin{equation}
V :\quad T \rightarrow -\frac{2T+1}{2T} \ .
\label{V}
\end{equation}
$W$ and $V$ generate the little groups of the fixed points $i/\sqrt 2$
and $(i-1)/2$ which are of order 2 and 4, respectively.
The fundamental domain is similar to the one of $SL(2,{\bf Z})$, defined by
\begin{equation}
\left\{  T_2>0, -{1\over 2}\le T_1\le 0, |T|\ge {1\over\sqrt{2}} \right\}\cup
\left\{  T_2>0, 0< T_1< 1/2, |T|> {1\over\sqrt{2}} \right\}
\label{funddom}
\end{equation}
with $T=T_1+iT_2$. The $SL(2,{\bf Z})$ fixed points $i$ and $e^{2i\pi\over 3}$
of order 2 and 3 are now replaced by $i/\sqrt 2$ and
$(i-1)/2$ of order 2 and 4, while $i\infty$ remains a cusp of width one.
The function $h$ of eq.(\ref{hfun}) is the analog of the $j$-invariant
function of $SL(2,{\bf Z})$. The numerator and denominator in (\ref{hfun}) are
respectively modular and cusp forms  of weight 8 for ${\Gamma_{0}(2)}_+$, so
that $h$ is of weight 0 and it has a pole of order one at $i\infty$ and a zero
of order $4$ at $(i-1)/2$. Moreover, $h$ is a bijection
between the fundamental domain (\ref{funddom}) and the Riemann sphere with
three punctures such that $h(i/\sqrt 2)=256$, $h((i-1)/2)=0$ and
$h(i\infty)=\infty$. Finally the solutions of the equation (\ref{h=256}) for
the conifold singularity in the perturbative limit consist in the images of
$i/\sqrt 2$ under the action of ${\Gamma_{0}(2)}_+$, which shows also that this
group is the modular one for $T$.

\section{The heterotic dual realization}

The dual of the above type II model should be a rank-3 $N=2$ heterotic vacuum
with one vector multiplet modulus $T$ besides the dilaton $S$, and 87
hypermultiplets. The method of constructing such models is described in
ref. \cite{Vafa}. The effective two derivative interactions of vector
multiplets
are described by a prepotential which has the general form of eq.(\ref{prep}).
The K\"{a}hler potential of the moduli metric can then be written as
\begin{eqnarray}
{ K} & = &  -\ln (i { Y})\qquad \mbox{ with }\label{potentiel} \\
{{Y}} & = & 2{ F}-2\bar{{ F}} - (T-\bar{T})
({ F}_{T}+\bar{{F}}_{\bar{T}})- (S-\bar{S})
({ F_{S}}+\bar{{ F}}_{\bar{S}}) \nonumber\\
& = & 2(S-{\bar S})(T-{\bar T})^2+2(f+f^{^{NP}}-
{\bar f}-{\bar f}^{^{NP}}) \nonumber\\
& &\qquad\qquad\qquad - (T-{\bar T})
(f_T+f^{^{NP}}_T+ {\bar f}_{\bar T}+{\bar f}^{^{NP}}_{\bar T})
-(S-{\bar S}) (f^{^{NP}}_S +{\bar f}^{^{NP}}_{\bar S}) \nonumber
\end{eqnarray}
where the subscripts denote differentiation with respect to the corresponding
fields. Note that $F$ is defined up to the addition of any quadratic polynomial
in $T$ and $S$ with real coefficients which leaves $Y$ invariant.

We now focus on the one loop contribution $f(T)$ which is actually the complete
perturbative correction due to $N=2$ non-renormalization theorems based on
analyticity and the invariance under the axionic shift. The one loop
correction to the K\"ahler metric is obtained by expanding
eq.(\ref{potentiel}),
\begin{equation}
{ K}_{T {\bar T}} = K^{(0)}_{T \bar{T}}\left[1+
\frac{2i}{S- \bar{S}}{\cal I} + \cdots\right]\ ,
\label{K_ttbar}
\end{equation}
where the tree-level metric $K^{(0)}_{T\bar{T}} =-2/(T-\bar{T})$ and
${\cal I}$ is given by
\begin{equation}
{\cal I}(T,{\bar T}) = \frac{i}{8} \left( \partial_T-\frac{2}{T-
\bar{T}} \right)
\left( \partial_T-\frac{4}{T- \bar{T}} \right) f(T) + c.c.
\label{I}
\end{equation}
${\cal I}$ can be computed by a one loop string calculation of an amplitude
involving the antisymmetric tensor using the method of ref.\cite{yuka}. From
the expression (\ref{I}) one can easily deduce the fifth derivative of the one
loop prepotential, $f^{(5)}$, \cite{W2g}:
\begin{equation}
f^{(5)}(T) = -\frac{8i}{(T- \bar{T})^{2}} \partial_{T}^{3} \left( (T-
\bar{T})^{2} {\cal I} \right)\ .
\label{f5}
\end{equation}

Since $T$ belongs to the coset $O(2,1)/O(2)$, the classical duality group is in
general a subgroup of $GL^{+}(2,{\bf{Z}})$ which is the set of fractional
linear
transformations of the form (\ref{Mt}) with $M$ being an integer matrix with
positive determinant. As $\cal I$ is related to a physical amplitude, it is
modular invariant. It is then straightforward to verify from eq.(\ref{f5}) that
$f^{(5)}$ is a modular function of weight 6.

Integrating equation (\ref{f5}) one can determine $f$ up to a quartic
polynomial,
\begin{equation}
f(T)= \int_{T_{0}}^{T} \frac{(T-T')^{4}}{4!} f^{(5)}(T') dT'\ ,
\label{forme integrale}
\end{equation}
where $T_0$ is an arbitrary point. The path of integration should not cross any
singularity of $f^{(5)}$, while the result of the integral depends on the
homology class of such paths. Different choices of homology classes of
paths change $f$ by quartic polynomials. Moreover under a modular
transformation (\ref{Mt}), $f$ does not transform covariantly. Using its
integral representation (\ref{forme integrale}), we see that it has a
weight $-4$ up to an addition of a quartic polynomial, ${\cal M}(T)$,
\begin{equation}
f(T) \rightarrow f(MT)=(\det M )^{2} (cT+d)^{-4} [f(T)+{\cal M}(T)]\ .
\label{transfo modulaire}
\end{equation}
These transformations should leave the physical metric (\ref{I}) invariant.
Hence, one must have
\begin{equation}
i\left( \partial_T-\frac{2}{T- \bar{T}} \right)
\left( \partial_T-\frac{4}{T- \bar{T}} \right) {\cal M}(T) + c.c. = 0,
\label{real}
\end{equation}
which is satisfied only if ${\cal M}(T)$ is a quartic polynomial with real
coefficients. In fact, we will see below that this
ambiguity is related to the non-trivial quantum monodromies \cite{AFGNT}.

Finally, in order to guarantee modular invariance of the full effective action,
the dilaton should also transform. Imposing the general condition that
modular transformations may be compensated by K\"ahler transformations in
eq.(\ref{potentiel}), one finds:
\begin{equation}
S \rightarrow  S -{c\over 2}\frac{f_T + {\cal M}_T}{cT+d} +c^2
\frac{f+{\cal M}}{(cT+d)^2} +\frac{1}{12}{\cal M}_{TT}+\lambda_M\, ,
\label{Stransf}
\end{equation}
up to an additive real constant $\lambda_M$ which corresponds to the
perturbative
heterotic symmetry of the axion shift. It follows that in the presence of
one loop
corrections one can define a $T$-duality invariant dilaton $S_{\rm inv}$
\cite{deWit},
\begin{equation}
S_{\rm inv}\equiv S-{1\over 12}f_{TT}\ ,
\label{Sinv}
\end{equation}
which however is not a special coordinate of $N=2$ K\"ahler geometry.

The above discussion applies to any rank-3 $N=2$ heterotic string
compactification. We now specialize to the dual candidate of the type II model
described in Section 2, for which the classical duality group is
${\Gamma_{0}(2)}_+$. Furthermore the order 2 fixed point in the fundamental
domain ({\ref{funddom}), $T=i/{\sqrt 2}$, which was found from the conifold
singularity in the type II model, should correspond in the heterotic
realization to the perturbative $SU(2)$ enhanced symmetry point. Using the
expressions for the left and right momenta, $p_L$ and $p_R$, of the
$\Gamma^{(2,1)}$ compactification lattice \cite{W2g},
\begin{eqnarray}
p_L &=& \frac{i\sqrt{2}}{T-\bar{T}} (n_1 + n_2 \bar{T}^2 + 2m
\bar{T} )
\nonumber \\
p_R &=& \frac{i\sqrt{2}}{T-\bar{T}} (n_1 + n_2 T\bar{T} +
m (T+\bar{T}) )\, ,
\label{mom}
\end{eqnarray}
one finds that the condition of additional massless states $p_L=0, p_R^2\le 2$
at $T=i/{\sqrt 2}$ is satisfied for two solutions,
\begin{equation}
SU(2)\ :\quad\quad n_1=\pm{1\over 2}\ ,\ n_2=\pm 1\ ,\ m=0\ ,
\label{states}
\end{equation}
for which $p_R=\pm 1$. This implies that the charges for the vector multiplets
sit in a lattice $\Gamma_{1,0}$ defined by $n_2$ being odd integer,
$n_1\in{\bf Z}+{1\over 2}$ and $m\in{\bf Z}$, so that the spectrum is invariant
under the modular group ${\Gamma_{0}(2)}_+$. Moreover, since $\Gamma_{1,0}$ is
odd, the enhanced $SU(2)$ symmetry at $T=i/{\sqrt 2}$ has a Kac-Moody
level 2 and a coupling constant $g=g_s/{\sqrt 2}$. Note that there is no other
point in the fundamental domain of $T$ with additional charged massless states.
However this lattice is not self-dual with respect to the inner product
$\frac{1}{2}(p_L\bar{p}_L'+\bar{p}_L p_L') - p_R p_R' = (2mm'-n_1 n_2' -n_2
n_1')$. World-sheet modular invariance under $\tau\rightarrow -1/\tau$ requires
the existence of additional vector multiplets contained in the dual lattice
which splits into 4 different sectors, with respect to their
transformations under
$\tau\rightarrow\tau +1$, labeled by $\Gamma_{1,\epsilon}$ and
$\Gamma_{2,\epsilon}$ for $\epsilon=0, {1\over 2}$. Applying once again the
transformation $\tau\rightarrow\-1/\tau$, for instance in
$\Gamma_{2,0}$, one finds 2 more sectors labeled by $\Gamma_{3,\epsilon}$:
\begin{equation}
\left.
\begin{array}{ccll}
\Gamma_{1,\epsilon} & : & n_1\in{\bf Z}+{1\over 2} & n_2\ {\rm odd} \\
\Gamma_{2,\epsilon} & : & n_1\in{\bf Z} & n_2\ {\rm even} \\
\Gamma_{3,\epsilon} & : & n_1\in{\bf Z}+{n_2+1\over 2} & n_2\in{\bf Z}
\end{array}
\right\} {\rm and}\quad m\in{\bf Z}+\epsilon
\end{equation}
These 6 sectors couple to different blocks of the remaining conformal field
theory in a way consistent with world-sheet modular invariance. At $T=(i-1)/2$,
one could get additional charged massless states in the sectors
$\Gamma_{1,{1\over 2}}$ and $\Gamma_{2,0}$ corresponding to
$n_1={n_2\over 2}=m=\pm 1$ and $\pm{1\over 2}$, respectively. However, the
corresponding coefficients associated to the remaining conformal field theory
should vanish since one knows that there are no extra charged massless states
at
this point. A similar argument applies for $T=i$ in the sector $\Gamma_{3,0}$
and $T=e^{2i\pi\over 3}$ in the sector
$\Gamma_{3,{1\over 2}}$ where one could have more charged massless states
corresponding to $n_1=n_2=\pm 1, m=0$ and $n_1=n_2=2m=\pm 1$, respectively.

Using the results of ref. \cite{yuka} it is straightforward to obtain an
expression for ${\cal I}$ as an integral over the complex Teichm\"uler
parameter
$\tau=\tau_1+i\tau_2$ of the world-sheet torus inside its fundamental domain,
\begin{equation}
{\cal I}  = \sum_{\ell=1}^6 \int\frac{d^2\tau}{\tau_{2}^{\; 3/2}}
{\bar C}_{\ell}(\bar{\tau})\, \partial_{\bar{\tau}}(\tau_2^{\; 1/2}
\sum_{p_L,p_R\in \Gamma_{\ell}}e^{\pi i\tau\, |p_{L}|^2}\,
e^{-\pi i \bar{\tau\,} p_{R}^2})   \ ,
\label{I1}
\end{equation}
where the sum extends over the six sectors denoted here by $\Gamma_{\ell}$, and
$\bar{C}_{\ell}$'s are $T$-independent modular functions with well-defined
transformation properties dictated by modular invariance of the integrand.
Actually $\bar{C}_{\ell}$ is the trace of
$(-1)^F q^{L_0-c/24} \bar{q}^{\bar{L}_0-\bar{c}/24}/{\bar\eta}$ in the Ramond
sectors of the corresponding remaining conformal blocks. We now use
eq.(\ref{f5}) together with the form of the lattice momenta (\ref{mom}) to
obtain:
\begin{equation}
f^{(5)}(T)= {512\pi^2 \over (T-{\bar T})^3} \sum_{\ell=1}^6
\int\frac{d^2\tau}{\tau_{2}^{\; 3/2}}\bar{C}_{\ell}(\bar{\tau})\,
\partial_{\bar{\tau}}\left[ \tau_2^2 \partial_{\bar{\tau}} (\tau_{2}^{\; 3/2}
\sum_{p_L,p_R\in \Gamma_{\ell}}p_L^3{p}_{R}\, e^{\pi i\tau\, |p_{L}|^2}\,
e^{-\pi i \bar{\tau}\, p_{R}^2})\right]\, .
\label{f51}
\end{equation}
One can easily show using the expression for the lattice momenta
(\ref{mom}) that
the integrand of $\partial_{\bar T}f^{(5)}$ reduces to a total derivative in
$\tau$ and the integral vanishes after integration by parts.

As $\tau_2\rightarrow\infty$, $2i\pi^2\bar{C}_{\ell}$ behaves as ${\bar
q}^{-1}$
and the one loop metric (\ref{I1}) has a logarithmic singularity at
$T=i/{\sqrt 2}$
associated to the massless states (\ref{states}),
\begin{equation}
{\cal I}\sim{1\over\pi}\ln \left| T-{i\over{\sqrt 2}}\right|^2
\qquad \mbox{ for }\ \ \  T \rightarrow i/\sqrt2\ .
\label{singul}
\end{equation}
This singularity can also be found in the effective field theory by taking into
account that $\left| T-{i\over{\sqrt 2}}\right|$ is proportional to the
mass of the two
$N=2$
vector multiplets near the $SU(2)$ symmetric point, as it can be seen from
eqs.(\ref{mom}) and (\ref{states}). Note that the coefficient of the
singularity is
half of the corresponding coefficient in the other known rank-3 dual-pair
example
\cite{{Vafa},{Louis},{W2g}} because in our case $SU(2)$ is realized with a
Kac-Moody level
2. The singularity (\ref{singul}) implies that the one loop prepotential
behaves
as
\begin{equation}
f\sim{4\over i\pi}\left(T-{i\over{\sqrt 2}}\right)^2 \ln
\left(T-{i\over{\sqrt 2}}\right)\ .
\label{fsingul}
\end{equation}

One can now use the duality symmetry ${\Gamma_{0}(2)}_+$ to determine $f$.
As mentioned above, its fifth derivative $f^{(5)}$ is a modular function of
weight 6. Moreover it has a triple pole at $T=i/{\sqrt 2}$ with coefficient
$16/i\pi$ to reproduce the logarithmic singularity (\ref{fsingul}), while it
vanishes at $T\rightarrow i\infty$, as seen from the integral representation
(\ref{f51}), implying that the metric does not grow at infinity. These
requirements
determine $f^{(5)}$ up to a numerical constant which can also be fixed by the
condition that the monodromy of $f$ as $T$ goes around $i/{\sqrt 2}$ is
contained
in the symplectic group $Sp(6,{\bf Z})$ as dictated by
$N=2$ supergravity. The result is:
\begin{equation}
f^{(5)}(T) = \frac{64}{i\pi} \left({h_T\over h-h\left({i\over{\sqrt
2}}\right)}\right)^3
{5h+3h\left({i\over{\sqrt 2}}\right)\over h^2} \ ,
\label{d5f}
\end{equation}
where the modular invariant function $h$ is given in eq.(\ref{hfun}).

A first non-trivial check of the proposed duality between this heterotic
model and
the type II compactification described in Section 2, is the comparison of
the two
corresponding prepotentials $F$ and ${\cal Y}$ of eqs.(\ref{prep}) and
(\ref{crochet}). Indeed, the identification $t_1=T$ and $t_2=S$ implies that
\begin{equation}
f^{(5)}=g^{(5)}=4\pi^2\sum_{n \geq 1}n^2 y_nq_{_{1}}^{n} \ ,
\end{equation}
which we verified up to the fifth order in the $q_{_1}$ expansion using the
numerical values for the coefficients $y_n$'s entering in the expression of the
Yukawa couplings (\ref{yukawa}) and given in ref. \cite{Candelas}.

\section{Perturbative duality group}

As we already mentioned, at the classical level the $T$-duality group of the
heterotic model is ${\Gamma_{0}(2)}_+$ generated by the transformations
$W$ and $V$ defined in eqs.(\ref{W}) - (\ref{V}). These generators obey the
relations
\begin{equation}
W^2=V^4=1 \qquad\qquad UVW=1 \ .
\label{group rel}
\end{equation}
Note the similarity of these relations with those of $SL(2,{\bf Z})$ obtained
by
replacing $V$ with a generator of order 3.

The order 2 generator $W$ corresponds to the Weyl reflection
of the $SU(2)$ gauge group at the enhanced symmetry point $T=i/{\sqrt 2}$
\cite{Giveon}. Away from this point, $SU(2)$ is spontaneously broken by the
vacuum expectation value, $a$, of a Higgs field along the flat direction of the
scalar potential, with the identification $a\propto (T-i/{\sqrt
2})/(T+i/{\sqrt 2})$
near the non-abelian point. Thus, the ${\bf Z}_2$ transformation $T\rightarrow
WT=-1/(2T)$ acts on $a$ as the parity $a\rightarrow -a$ which remains unbroken
after the Higgs phenomenon.

At the quantum level, because of the singularity at $T=i/{\sqrt 2}$ the first
relation of (\ref{group rel}) is modified. $W^2$ instead of being the
identity, is
determined by the corresponding non-trivial monodromy $M_{i/\sqrt 2}$~,
\begin{equation}
W^2=M_{i/\sqrt 2}\qquad V^4=1 \qquad\qquad UVW=1 \ .
\label{group qrel}
\end{equation}
In fact, when moving a point around the singularity the one loop
prepotential $f$
transforms as:
\begin{eqnarray}
M_{i/\sqrt 2}\ :\quad &f(T)& \rightarrow  f(T) + {\cal M}_{i/\sqrt 2}(T)  \\
{\rm with}\qquad {\cal M}_{i/\sqrt 2}(T) &=&
-4 \left( T-{i\over{\sqrt 2}} \right)^{2} \left( T+{i\over{\sqrt 2}} \right)^2
\nonumber\\
&=& -4T^{4}-4T^{2}-1 \nonumber\ .
\nonumber
\end{eqnarray}
The polynomial ${\cal M}_{i/\sqrt 2}$ can be determined by expanding the
expression
(\ref{d5f}) around $T=i/{\sqrt 2}$. It can also be easily obtained
just from the knowledge of the leading behavior of $f$ in eq.(\ref{fsingul})
together with the reality condition on the coefficients of ${\cal
M}_{i/\sqrt 2}$
which is required for modular invariance of the physical quantity ${\cal
I}$ (see
eq.(\ref{real})).

One can now use the modified group relations (\ref{group qrel}) to determine
the
transformation properties of $f$ under the generators $W$ and $V$. This
amounts to
determine the corresponding quartic polynomials ${\cal M}$ entering in
eq.(\ref{transfo modulaire}), which we will denote by ${\cal M}_W$ and
${\cal M}_V$ respectively. The first relation of eq.(\ref{group qrel}) gives:
\begin{equation}
{\cal M}_W(T)+4T^4{\cal M}_W\left({-1\over 2T}\right)={\cal M}_{i/\sqrt 2}(T)\
{}.
\end{equation}
The general solution depends on two real parameters $w_0$ and $w_1$,
\begin{equation}
{\cal M}_W(T)= -4(1+w_{0})T^{4}+2w_{1}T^{3}-2T^{2}+w_{1}T+w_{0} \ .
\label{Wq}
\end{equation}
Imposing the second relation of eq.(\ref{group qrel}), one finds just one
linear
constraint among the five real coefficients $v_n$ of the polynomial ${\cal
M}_V$. Finally, using the third relation of eq.(\ref{group qrel}), one can
obtain
the polynomial ${\cal M}_U$ associated to the translation $U$, as a
function of the
six independent parameters $w_0$, $w_1$ and $v_n$,
\begin{equation}
{\cal M}_U(T) = -{\cal M}_W(T+1) -4(T+1)^4{\cal M}_V\left({-1\over
2T+2}\right)\ .
\label{Uq}
\end{equation}
We now note that there is an additional constraint to ${\cal M}_U$ coming from
the behavior of $f$ at infinity. Since $f^{(5)}$ vanishes as $T\rightarrow
i\infty$, $f$ behaves at most as $T^4$ which implies that the coefficient of
$T^4$ in ${\cal M}_U$ is zero. One then finds $v_0=1+w_0$ and the polynomial
${\cal M}_V$ becomes:
\begin{eqnarray}
&{\cal M}_V(T)& = \sum_{n=0}^4 v_n T^n
\label{Vq}\\
{\rm with} \quad \ & v_0=1+w_0 & \quad ; \quad
v_4=v_3-{4\over 3}v_2+2v_1-4v_0 \ .
\nonumber
\end{eqnarray}
We recall that the above transformations of $f$ are accompanied by suitable
$T$-dependent shifts of the dilaton, determined from eq.(\ref{Stransf}) up to
additive $T$-independent constants.

The five independent parameters $w_{0,1}$ and $v_{1,2,3}$ entering in the
polynomials ${\cal M}_W$ and  ${\cal M}_V$ can be chosen arbitrarily using the
freedom to add to $f$ a quartic polynomial with real coefficients. This
freedom is
valid only if one neglects the non-perturbative corrections, $f^{^{NP}}=0$ in
eq.(\ref{prep}), and it can be easily seen from eq.(\ref{potentiel}). More
precisely, adding to $f$ a function ${\cal P}(T)$ one finds that $Y$ remains
invariant provided ${\cal P}$ satisfies eq.(\ref{real}) which implies that
${\cal
P}$ is a quartic polynomial with real coefficients, and $S$ is redefined as
$S\rightarrow S+{\cal P}_{TT}/12$. In the presence of non-perturbative
corrections,
only the addition of a quadratic polynomial is allowed, which leaves $S$ inert,
corresponding to a change of basis.

Following the analysis of ref. \cite{AFGNT}, the perturbative $T$-duality group
$G$ contains a normal abelian subgroup $H$ which is generated by elements
obtained by conjugating the monodromy $M_{i/{\sqrt 2}}$ with elements
(\ref{Mt})
of ${\Gamma_{0}(2)}_+$. A general element of $H$ is then obtained by a sequence
of
such transformations, and shifts $f$ by:
\begin{equation}
f\rightarrow f +\sum_i N_i (\det M_i)^{-2}[2(a_iT+b_i)^2+(c_iT+d_i)^2]^2
\equiv f +\sum_{n=0}^4 c_nT^n \quad N_i\in{\bf Z} \ .
\label{H}
\end{equation}
Since the generated coefficients $c_n$'s are 5 independent integer
parameters, $H$
is isomorphic to ${\bf Z}^5$. Moreover, the quotient $G/H$ is isomorphic to
${\Gamma_{0}(2)}_+$ under which the $c_n$'s transform in the
{\underline{\bf 5}} representation (second rank traceless symmetric tensor)
\cite{AFGNT}. Thus, the perturbative $T$-duality group $G$ involves 8 integer
parameters. Note that $G$ is not a semidirect product of ${\Gamma_{0}(2)}_+$
and
$H$ because ${\Gamma_{0}(2)}_+$ is not a subgroup of $G$. Finally, the full
perturbative
symmetry group is the direct product of $G$ with the constant dilaton shift,
\begin{equation}
D:\quad S\rightarrow S+\lambda \ .
\label{D}
\end{equation}

The symplectic structure of $N=2$ supergravity implies that all symmetry
transformations of the effective low energy theory must be contained in the
symplectic group $Sp(6,{\bf R})$ \cite{{sugra},{Ceresole}} which is broken to
$Sp(6,{\bf Z})$ by
quantum effects. It is then convenient to introduce a field basis where all
transformations act linearly. As usual, we define
\begin{equation}
T=\frac{X^{1}}{X^{0}}\qquad \mbox{ and }\qquad S=\frac{X^{2}}{X^{0}}\ ,
\label{hombas}
\end{equation}
in terms of the homogeneous coordinates $X^I$ with $I=0,1,2$. The
prepotential is
a homogeneous function of degree 2,
\begin{equation}
F(X^{0},X^{1},X^{2})=(X^{0})^{2}
F\left(\frac{X^{1}}{X^{0}},\frac{X^{2}}{X^{0}}\right)\ ,
\end{equation}
and the K\"{a}hler potential (\ref{potentiel}) is given in terms of
\begin{equation}
Y = \bar{X}^{I} F_{I}-X^{I} {\bar F}_{I}\qquad {\rm with}\qquad
F_{I}=\frac{\partial F}{\partial X^{I}}\ .
\label{potentiell}
\end{equation}
In this way all symmetries must act in the basis $(F_I,X^I)$ as symplectic
transformations which leave the K\"{a}hler potential (\ref{potentiell})
manifestly
invariant. Their symplectic action on the homogeneous basis is uniquely
defined from the corresponding transformations of the fields $T,S$ and the
prepotential $F$.\footnote{There is an overall sign ambiguity which is
irrelevant for our purposes since we identify the element $-{\bf 1}$ with
the identity.}

The invariance of the action under symplectic transformations implies that
electric and magnetic charges form a symplectic vector. It is then useful
to use a
slightly different basis from (\ref{hombas}), in which the perturbative
transformations can be restricted consistently on the subspace of magnetically
neutral states and therefore do not involve the inversion of the string
coupling.
They should therefore be represented by matrices of the form
\begin{equation}
\left( \begin{array}{cl}
	{\bf a} & {\bf a{b}} \\
	{\bf 0} & {\bf a}^{\mit -1,t}
       \end{array}    \right)
\label{matrices per}
\end{equation}
where ${\bf {b}}$ is a $3\times 3$ symmetric matrix, ${\bf {b}}
= {\bf {b}}^{t}$. For ${\bf {b}}\ne 0$ the effective action
changes with $F{\tilde F}$ total derivative terms. Hence, classical duality
transformations are block diagonal while the perturbative monodromies generate
upper off-diagonal elements \cite{Ceresole}. The new basis, obtained by a
symplectic change of (\ref{hombas}), is
\begin{equation}
{\widetilde F}_I = F_{I}\ \ \widetilde{X}^{I}=X^{I}\ (I=0,1)\quad
\widetilde{F}_{2}=X^{2}\ \ \widetilde{X}^{2}=-F_{2} \ .
\label{newbasis}
\end{equation}

It is now straightforward to obtain the matrix representation of the
generators of
the perturbative duality group, using eq.(\ref{transfo modulaire}) with
(\ref{Wq}) - (\ref{Vq}). We first fix the constant dilaton shift $\lambda_M$
entering in the transformation of $S$ in eq.(\ref{Stransf}) by imposing the
group relations (\ref{group qrel}) $V^4=1$, and $UVW=1$ {\it i.e.} that $S$
is inert under the action of $V^4$ and $UVW$. The result is:
\begin{equation}
\lambda_V=0\qquad ; \qquad \lambda_U=w_1-v_1-\lambda_W \ .
\end{equation}
Note that $S$ is never inert under the action of the monodromy $W^2=M_{i/{\sqrt
2}}$ which corresponds to the first group relation. However, one can choose
$\lambda_W$ in a way that this monodromy coincides with the one obtained in
the rigid field theory as we will discuss below \cite{Lust}. In our case, this
requirement gives $\lambda_W=-{2/3}$. Furthermore, to simplify the expressions
one has to make a choice for the five parameters $w_{0,1}$ and $v_{1,2,3}$
entering in the polynomials ${\cal M}_W$ and
${\cal M}_V$. We first choose $w_0=-1/2$ and $w_1=0$, so that ${\cal M}_W$
equals half of the monodromy polynomial ${\cal M}_{i/{\sqrt 2}}$. We then
impose
the requirement that $S$ is inert under the translation $U$
which yields $v_1=w_1=0$ and $v_2=2-6\lambda_W=6$. Finally we choose $v_3=16$,
and the 3 polynomials (\ref{Wq}) - (\ref{Vq}) become:
\begin{eqnarray}
{\cal M}_W &=& -2T^4-2T^2-{1\over 2} \nonumber\\
{\cal M}_V &=& 6T^4+16T^3+6T^2+{1\over 2} \label{polynomes} \\
{\cal M}_U &=& -4T^2+3 \ .\nonumber
\end{eqnarray}
As a result, in the perturbative basis (\ref{newbasis}) the generators are:
\begin{eqnarray}
\widetilde{W} &=&	\left(	\begin{array}{cccccc}
		0&0&-1&1&0&1 \\
		0&1&0&0&0&0 \\
		-1&0&0&1&0&1 \\
		0&0&0&0&0&-1 \\
		0&0&0&0&1&0 \\
		0&0&0&-1&0&0
		\end{array}	\right)\quad
\widetilde{V} =	\left(	\begin{array}{cccccc}
		-2&1&-1&-3&0&3 \\
		-4&1&0&-4&8&4 \\
		-1&0&0&-1&0&-1 \\
		0&0&0&0&0&-1 \\
		0&0&0&0&1&1 \\
		0&0&0&-1&-4&-2
		\end{array}	\right)\label{matrices}\\
\widetilde{U}&=&(\widetilde{V}\widetilde{W})^{-1} =	\left(
\begin{array}{cccccc}
		1&-1&2&6&8&0 \\
		0&1&-4&0&-8&0 \\
		0&0&1&0&0&0 \\
		0&0&0&1&0&0 \\
		0&0&0&1&1&0 \\
		0&0&0&2&4&1
		\end{array}	\right)\quad
\widetilde{D} =	\left(	\begin{array}{cccccc}
		1&0&0&0&0&\lambda \\
		0&1&0&0&-4\lambda&0 \\
		0&0&1&\lambda&0&0 \\
		0&0&0&1&0&0 \\
		0&0&0&0&1&0 \\
		0&0&0&0&0&1
		\end{array}	\right)\nonumber
\end{eqnarray}
where in the second line we also give the matrix representation of the dilaton
shift (\ref{D}). The upper off-diagonal $3\times 3$
matrix is in fact $\lambda$ times the $O(2,1)$ metric in the perturbative basis
$(1,T,2T^2)$ we are using \cite{{Ceresole},{AFGNT}}. The parameter $\lambda$
should also be quantized at the non perturbative level, $\lambda=1$, due to
instanton effects.

The perturbative monodromy group is generated by the 3 generators
$\widetilde{W}$,
$\widetilde{V}$ and $\widetilde{D}$. The generator $\widetilde{W}$, whose
classical
part corresponds to the Weyl reflection of the $SU(2)$ enhanced gauge
symmetry as
we discussed in the beginning of this section, coincides with the perturbative
monodromy $M_{\infty}$ of the rigid theory \cite{Seiberg-Witten}. In fact by a
linear change of our field variables which diagonalizes its classical part,
\begin{equation}
{\widehat F}_0={{\widetilde F}_0+{\widetilde F}_2\over{\sqrt 2}}\ ,\
{\widehat F}_1={\widetilde F}_1\ ,\
{\widehat F}_2={{\widetilde F}_0-{\widetilde F}_2\over{\sqrt 2}}\ ,\
{\widehat X}^0={{\widetilde X}^0+{\widetilde X}^2\over{\sqrt 2}}\ ,\
{\widehat X}^1={\widetilde X}^1\ ,\
{\widehat X}^2={{\widetilde X}^0-{\widetilde X}^2\over{\sqrt 2}}\ ,
\label{rigidbasis}
\end{equation}
$\widetilde{W}$ takes the form:
\begin{equation}
\widehat{W} = {\widehat M}_{\infty}=	\left(	\begin{array}{cccccc}
		-1&0&0&2&0&0 \\
		0&1&0&0&0&0 \\
		0&0&1&0&0&0 \\
		0&0&0&-1&0&0 \\
		0&0&0&0&1&0 \\
		0&0&0&0&0&1
		\end{array}	\right)
\end{equation}
Near the enhanced symmetry point, ${\widehat X}^0\sim 2i(T-i/{\sqrt 2})$
and the
subspace $({\widehat F}_0,{\widehat X}^0)$ is associated to the rigid
supersymmetric theory. The generator $\widehat{W}$ acts non-trivially only
on this
subspace and its action represented by the corresponding $2\times 2$
submatrix can
be identified with $M_{\infty}$ of ref. \cite{Seiberg-Witten}.

\section{Exact symmetry group}

The exact symmetry group of this model can be determined using duality from its
type II realization based on the Calabi-Yau manifold $X_8$, discussed in
Section
2. The full monodromy group was worked out in ref. \cite{Candelas} and was
shown
to be a subgroup of $Sp(6,{\bf Z})$, ${\cal G}$, generated by 3 elements
denoted by
$A$, $T$ and $B$.\footnote{To keep the same notation as ref. \cite{Candelas},
we
are forced to use the symbol $T$ for the generator which should not be confused
with the heterotic modulus.} The element $A$ generates an exact ${\bf Z}_8$
symmetry defined in eq.(\ref{ordre 8}) and satisfies $A^8=1$. The other two
generators $T$ and $B$ are associated to the monodromies around the conifold
and
the strong coupling loci, respectively, described by the discriminant
(\ref{conifold}), and they are subject to some group relations given in ref.
\cite{Candelas}. Moreover, by considering the large complex structure limit,
two
independent (mutually commuting) translations were identified, acting on the
special coordinates
$t_{1,2}$ of eq.(\ref{mirror map}),
\begin{eqnarray}
\label{tshifts}
S_1 &=& (AT)^{-2}:\quad t_1\rightarrow t_1+1  \nonumber\\
S_2 &=& (ATB)^{-1}:\quad t_2\rightarrow t_2+1 \ .
\end{eqnarray}

Our first task is to identify the generators of the perturbative heterotic
duality group derived in the previous section as elements of the type II
monodromy
group ${\cal G}$. {}From the identification with the heterotic variables
$t_1=T$
and $t_2=S$, one concludes that in a suitable basis one should have:
\begin{equation}
S_1=U\qquad S_2=D|_{\lambda =1} \ ,
\label{s12}
\end{equation}
where the matrix representation of $U$ and $D$ is given in
eq.(\ref{matrices}), in
the perturbative basis (\ref{newbasis}). The one loop heterotic
prepotential $f$ is shifted by the polynomial ${\cal M}_U$ of eq.
(\ref{polynomes})
under the action of $U$, while it remains inert under $D$. These
transformations
allow us to determine the integration constants entering in the expression
of the
type II prepotential (\ref{crochet}) for which the identification (\ref{s12})
should be valid,
\begin{equation}
\alpha=4 \qquad \delta=3 \qquad \beta=\gamma=\epsilon=0\ .
\label{choix}
\end{equation}
On the other hand, the 3 generators $A,T,B$ are given in ref. \cite{Candelas}
in
the symplectic basis $(F_I,X^I)$ of eq.(\ref{hombas}) for a different choice of
these parameters, namely $\alpha=\gamma=0$, $\beta=-2$, $\delta=3$ and
$\epsilon=1$.\footnote{Note the sign mistakes in the first equation of page
521 of ref. \cite{Candelas} where one reads ``$\delta=-3$, $\epsilon=-1$"
instead
of their opposite values which lead to the correct conjugation matrix $m$.} The
values of the parameters can be changed by a conjugation with the symplectic
matrix
\begin{equation}
\left( \begin{array}{cc}
        {\bf 1} & Q \\
        {\bf 0} & {\bf 1}
       \end{array} \right)\quad \mbox{with}\quad Q = \left( \begin{array}{ccc}
							0 & \delta' & \epsilon' \\
							\delta '& \alpha' & \beta' \\
							\epsilon' & \beta'  &\gamma'
						\end{array} \right)
\end{equation}
under which $\alpha, \beta,\cdots, \epsilon$ are shifted by
$\alpha',\beta', \cdots,
\epsilon'$. Performing such a conjugation, the two translations are identified
according to eq.(\ref{s12}) in the basis (\ref{choix}), and the generators
$A,T,B$
are:
\begin{eqnarray}
A &=&	\left(	\begin{array}{cccccc}
		-2&0&1&-2&0&-1 \\
		-4&1&0&-4&4&0 \\
		-3&1&-1&-4&0&-1 \\
		1&0&0&1&0&1 \\
		-1&0&0&-1&1&0 \\
		1&0&0&1&0&0
		\end{array}	\right)\quad
B =	\left(	\begin{array}{cccccc}
		1&-1&2&-2&0&4 \\
		0&1&0&0&0&0 \\
		0&1&-1&0&0&-2 \\
		0&0&0&1&0&0 \\
		0&0&0&-1&1&1 \\
		0&0&0&2&0&-1
		\end{array}	\right)\label{ABT}\\
T &=&	\left(	\begin{array}{cccccc}
		1&0&0&0&0&0 \\
		0&1&0&0&0&0 \\
		1&0&1&0&0&1 \\
		-1&0&0&1&0&-1 \\
		0&0&0&0&1&0 \\
		0&0&0&0&0&1
		\end{array}	\right)\nonumber
\end{eqnarray}

The fact that the generator $A$ is of order 8 suggests that the order 4
generator
$V$ should be identified with a conjugation of $A^2$ (or its inverse). Indeed
it
is easy to verify that the matrix $\widetilde{V}$ transformed in the original
basis (\ref{hombas}) is identical to $A^2$. Thus, we have shown that the
perturbative dualities generated by $U$ and $V$, together with the quantized
dilaton shift $D|_{\lambda=1}$, form a subgroup of the type II symmetries.
Using eqs. (\ref{tshifts}), (\ref{s12}) and the group relations (\ref{group
qrel})
one has:
\begin{equation}
U=(AT)^{-2}\ ,\quad V=A^2 \ ,\quad W=A^{-1}TAT\ ,\quad
D|_{\lambda=1}=(ATB)^{-1}\ .
\label{pertgr}
\end{equation}

Our next task is to identify the quantum monodromy group $\Gamma(2)$ of the
$SU(2)$ rigid field theory as a subgroup of ${\cal G}$. It is generated by two
elements, $M_{\infty}$ and $M_1$, which satisfy the relation
\cite{Seiberg-Witten}:
\begin{equation}
M_{\infty}=M_1M_{-1}\ ,
\label{gamma2}
\end{equation}
where $M_{\infty}$ is the perturbative monodromy, while $M_1$ and $M_{-1}$
correspond to the monodromies around the points where dyonic hypermultiplets
become massless and they are conjugate to each other. These properties can be
used as a guide for the identification \cite{Lust}. We have shown in the
previous section that $M_{\infty}$ coincides with the generator $W$ of the
heterotic duality group. A simple inspection of its form (\ref{pertgr})
suggests
that $M_1$ should be identified with $T$ (or its conjugate $A^{-1}TA$). Indeed
one can easily verify that in the basis (\ref{rigidbasis}) the generator $T$
takes the form:
\begin{equation}
{\widehat T} ={\widehat M}_1=\left(	\begin{array}{cccccc}
		1&0&0&0&0&0 \\
		0&1&0&0&0&0 \\
		0&0&1&0&0&0 \\
		-2&0&0&1&0&0 \\
		0&0&0&0&1&0 \\
		0&0&0&0&0&1
		\end{array}	\right)
\end{equation}
One sees that in analogy with ${\widehat W}$,  ${\widehat T}$ acts
non-trivially
only on the subspace of the rigid supersymmetric theory $({\widehat
F}_0,{\widehat
X}^0)$, and its action represented by the corresponding $2\times 2$ submatrix
coincides with $M_1$ of ref. \cite{Seiberg-Witten}. Thus, we have shown that
the
non-perturbative monodromies of the $SU(2)$ rigid field theory form a
$\Gamma(2)$
subgroup of the type II symmetry ${\cal G}$,
\begin{equation}
M_{\infty}=W\ ,\quad M_1=T\ ,\quad M_{-1}=(AT)^{-1}T(AT)\ .
\label{rigidgr}
\end{equation}

This result provides a non-perturbative test of the string duality and
confirms that
the $SU(2)$ enhanced symmetry which is present in perturbation theory of $N=2$
heterotic compactifications is never restored when non-perturbative effects are
taken into account (except in the trivial case of zero coupling). Instead,
there
are two singular (complex) lines in the moduli space of $S$ and $T$ where
massless
dyonic hypermultiplets appear with charges determined in the rigid theory.
String
duality maps these lines into the conifold locus of the Calabi-Yau manifold
while
the dyonic hypermultiplets are exchanged with the solitonic back hole
hypermultiplets of type II string \cite{Stro}. The position of the singular
lines
is determined by the equation $x=2^{-8}/\left(1\pm 2{\sqrt y}\right)$ in
terms of the large
complex structure variables $x$ and $y$ of eq.(\ref{mirror map}).

The heterotic $T$-duality group extended with the quantized dilaton shift
and the
non-perturbative monodromy of the rigid field theory (\ref{pertgr}),
(\ref{rigidgr}) form a subgroup of the full exact symmetry ${\cal G}$
generated,
for instance, by the elements $A^2$, $A^{-1}TA$, $A^{-1}B$, and
$T$. One needs just to introduce the ``square root" of the perturbative
generator
$V=A^2$, or equivalently $B$, to recover the full ${\cal G}$. The generator $B$
is related to the monodromy around the non-perturbative singularity $y=1/4$
corresponding to infinite coupling, where new massless dyonic
hypermultiplets appear with charges under the ``dilaton" $U(1)$. This singular
line which is not present in the rigid theory is a new stringy phenomenon
related
to the dilaton and seems to be a generic feature of string vacua. It
would be interesting to find the corresponding solitonic solutions which
become massless in the type II theory, in analogy with the black holes of the
conifold locus.

{}From the matrix representation of the generator $B$ in eq.(\ref{ABT}), one
finds
that it corresponds to the transformation,
\begin{equation}
B: \qquad \begin{array}{ccl}
T&\rightarrow&T+S+1\\
S&\rightarrow&-S+2\\
F(T,S)&\rightarrow&F(T,S)+S^2-1
\end{array}
\end{equation}
where $F$ is the full prepotential (\ref{prep}). The monodromy polynomial
$S^2-1$
follows from the upper $3\times 3$ off-diagonal block of the matrix.
Note that redefining the dilaton as $S'\equiv S+T-1$, $B$ corresponds to the
exchange $T\leftrightarrow S'$ as suggested in ref. \cite{Klemm.95}. In fact
$B^2$ corresponds to the monodromy one obtains when moving a point around the
(ultra-)strong coupling singularity $S=1$,
\begin{equation}
B^2:\qquad F\rightarrow F+2(S-1)^2
\end{equation}
which suggests that $F$ has a logarithmic singularity at $S=1$ (modulo integer
shifts by $S_2$).

One would like to use the exact symmetries discussed above in order to
determine
the low energy effective action of vector multiplets based on the
prepotential $F$.
Consider for instance $\partial_T^3 F$ which is invariant under both
transformations $B$ and $S_2$ (or equivalently $AT$). Unfortunately, to
determine
this function one needs an information about one more generator, $T$ or $A$
whose
action on our variables is highly non-trivial. Similarly, the perturbative
transformation $W$ (or $V$) is drastically modified.

\vspace{1.cm}
{\bf Acknowledgments}
We would like to thank J. Lascoux and K.S. Narain for useful discussions.

\end{document}